\documentclass{article}

\usepackage{arxiv}

\usepackage[utf8]{inputenc} 
\usepackage[T1]{fontenc}    
\usepackage{hyperref}       
\usepackage{url}            
\usepackage{booktabs}       
\usepackage{amsfonts}       
\usepackage{nicefrac}       
\usepackage{microtype}      
\usepackage{lipsum}		
\usepackage{graphicx}
\usepackage{doi}

\title{On the Local equivalence of the Black Scholes and the Merton Garman equations}

\author{
  Ivan Arraut \\
 University of Saint Joseph\\
  Estrada Marginal da Ilha Verde, 14-17, Macao, China\\
  \texttt{ivan.arraut@usj.edu.mo}} 

\begin{document}
\maketitle

\begin{abstract}
It was demonstrated previously that the stochastic volatility emerges as the gauge field necessary for restoring the local symmetry under changes of the prices of the stocks inside the Black-Scholes (BS) equation. When this occurs, then a Merton-Garman-like equation emerges. From the perspective of manifolds, this means that the Black-Scholes equation and the Merton-Garman (MG) one can be considered as locally equivalent. In this scenario, the MG Hamiltonian is a special case of a more general Hamiltonian, here called gauge-Hamiltonian. We then show that the gauge character of the volatility implies some specific functional relation between the prices of the stock and the volatility. The connection between the prices of the stocks and the volatility, is a powerful tool for improving the volatility estimations in the stock market, which is a key ingredient for the investors to make good decisions. Finally, we define an extended version of the martingale condition, defined for the gauge-Hamiltonian.
\end{abstract}

\keywords{Merton-Garman equation \and Black-Scholes equation \and Gauge theory}

\section{Introduction}

The Black Scholes (BS) equation emerged as the first equation able to make predictions for the prices of the Options inside the stock market \cite{BS}. The authors of this equation brilliantly generated a risk-free portfolio by combining the option prices with their derivatives. Inside the BS equation, the volatility is a free-parameter which has to be estimated by the investors \cite{Vol}. The investors usually compare their estimations with the historical value of the volatility (from the charts) in order to decide whether or not to buy some specific option. After the BS equation, the Merton-Garman one was developed independently, this time considering the volatility as a stochastic variable in the same way as we deal with the prices of the stocks \cite{Op3, Merton2}. For a long time there was no fundamental connection between both equations. However, understanding that the MG equation should be a natural extension of the BS case, then it is normal to suspect that a fundamental principle able to connect both equations must exist. In \cite{Equivalence, Ivan}, it was demonstrated, by using the gauge principle, that there exist a natural connection between the MG and the BS equations.
If we impose symmetry under local changes on the stock prices over the BS equation expressed in its Hamiltonian form, then the only way to maintain the symmetry is by introducing a gauge field, which transforms in a specific way for compensating any change generated by the local changes on the stock prices. Interestingly, it was discovered that this gauge field corresponds to the stochastic volatility. In other words, we obtain the MG equation by imposing local symmetry under changes of the prices of the stocks on the BS equation. This amazing discover means that the BS and the MG equations are locally equivalent. This also implies some potential applications of the gauge principle over the option market.
In this way, we can perceive the BS equation as the one living over a manifold where the volatility does not exist (basically is fixed arbitrarily), while the MG equation lives over a manifold deformed for the presence of the stochastic volatility. However, the volatility vanishes locally, and then the MG equations converges locally to the BS equation. The gauge principle used for analyzing this scenario, is general and it has been used on several research areas besides Quantum Finance \cite{Gauge1, Gauge2, physics1, physics2, physics3, physics4, physics5, Mypapers}. During the derivation of the MG equation from the BS Hamiltonian, strong constraints over the free-parameters of the gauge Hamiltonian in order to match with the MG one. In this paper we review the local equivalence of the BS and MG equations. We then consider the general gauge Hamiltonian obtained in \cite{Equivalence}, without imposing any restriction over the free-parameters of the system or over the stochastic volatility. Finally, we focus on the martingale state of the system in order to find relations between the prices of the stock and the stochastic volatility. These new relations can be used in future for doing estimations about the possible values taken by the volatility of some specific stock. Some preliminary relations, but considering some constraints, were considered in \cite{Ivan, Equivalence}. We argue that the estimations on the possible values taken by the stochastic volatility can be improved by using the gauge principle proposed in \cite{Equivalence}. This is possible because the gauge principle suggests that the variations on the prices of the stocks are not independent on the changes in volatility but rather, there is a well established functional relation between both variables. Inside the knowledge of the author, a clear connection between both variables in the way suggested by the gauge principle, has not been proposed until now. This is the main contribution of the paper. The paper is organized as follows: In Sec. (\ref{First sectio}), we revise the standard formulation of the BS equation, together with its Hamiltonian formulation. In Sec. (\ref{Merton}), we revise the standard formulation of the Merton-Garman equation, together with its Hamiltonian formulation. In Sec. (\ref{Symmetr...}), we develop the gauge version of the BS equation, which comes out to be equivalent to a MG-like equation by deriving the most general gauge Hamiltonian. We then explain under which conditions the MG and the BS equations can be equivalent locally. We extend the analysis for including the gauge-Hamiltonian. In Sec. (\ref{Martingalelalala}), we use the relation gauge relation between the volatility and the stock prices for analyzing the functional dependence between both variables, with the purpose of improving the volatility estimations. Finally, in Sec. (\ref{Conclusions}), we conclude. 

\section{The Black Scholes equation}   \label{First sectio}

If we define the stock price as $S(t)$, which is normally taken as a random stochastic variable, then its evolution is defined in agreement with the equation \cite{B} 

\begin{equation}   \label{rainbowl}
\frac{dS(t)}{dt}=\phi S(t)+\sigma S(t)R(t).
\end{equation}   
In this equation $\phi$ is the expected return of the security, $R(t)$ is the Gaussian white noise with zero mean and $\sigma$ is the volatility. It is important to remark that here the volatility $\sigma$ is just a free-parameter of the system. In eq. (\ref{rainbowl}) it appears multiplying the Gaussian white noise. We cannot avoid random fluctuations for the stock market. The remarkable contribution of Black and Scholes, was the creation of a portfolio free of these fluctuations. This is achieved after combining the price of an Option with its derivative inside the portfolio such that the random fluctuations of the prices of the options are cancelled by some correlated random fluctuations coming from the derivative of the same option. The portfolio is then defined as

\begin{equation}   \label{forever}
\Pi=C-\frac{\partial C}{\partial S}S.
\end{equation}
Since this portfolio is free of any random fluctuation, then we can define its dynamics and then its derivative. Its evolution can be predicted by using standard techniques. Eq. (\ref{forever}) is a portfolio where an investor holds the Option and then {\it short sells} the amount $\frac{\partial \psi}{\partial S}$ for the security $S$. From the Ito calculus (stochastic calculus) \cite{Stockbook}, we can calculate

\begin{equation}   \label{BS}
\frac{d\Pi}{dt}=\frac{\partial C}{\partial t}+\frac{1}{2}\sigma^2S^2\frac{\partial^2 C}{\partial S^2}.
\end{equation}
For a risk free-portfolio, as it is the case here, we can connect the total derivative, evaluated with respect to the time, with the interest rate $r$ as follows 

\begin{equation}   \label{hedged}
\frac{d\Pi}{dt}=r\Pi.
\end{equation} 
The combination of the results (\ref{forever}) and (\ref{BS}) brings out

\begin{equation}   \label{BSeq}
\frac{\partial C}{\partial t}+rS\frac{\partial C}{\partial S}+\frac{1}{2}\sigma^2S^2\frac{\partial^2C}{\partial S^2}=rC.
\end{equation}
This is the BS equation \cite{Op3, Merton2, B}, which is independent of the expectations of the investors $\phi$. Here we repeat the Basic assumptions of the Black-Scholes equation as follows:\\
1). The spot interest rate $r$ is constant.\\
2). In order to create the hedged portfolio $\Pi$, the stock is infinitely divisible, and in addition it is possible to short sell the stock.\\
3). The portfolio satisfies the no-arbitrage condition.\\
4). The portfolio $\Pi$ can be re-balanced continuously.\\
5). There is no fee for transaction.\\ 
6). The stock price has a continuous evolution. 

\subsection{Black-Scholes Hamiltonian formulation}

We can convert eq. (\ref{BSeq}) by doing a change of variable. The final equation is equal to the Schr\"odinger equation but this time with a non-Hermitian Hamiltonian \cite{Schro}. 
The process for deriving the BS Hamiltonian starts by considering the change of variable $S=e^x$, where $-\infty<x<\infty$. Then the BS equation becomes

\begin{equation}   \label{BSHamiltonian2}
\frac{\partial C}{\partial t}=\hat{H}_{BS}C,
\end{equation} 
where we have defined the operator  

\begin{equation}   \label{BSHamiltonian}
\hat{H}_{BS}=-\frac{\sigma^2}{2}\frac{\partial^2}{\partial x^2}+\left(\frac{1}{2}\sigma^2-r\right)\frac{\partial}{\partial x}+r,
\end{equation}
as the BS Hamiltonian which is not Hermitian. The non-Hermiticity of the Hamiltonian, can be perceived from the fact that the derivative with respect to the time of the Option price in eq. (\ref{BSHamiltonian}). A deeper proof of this statement was done in \cite{B}. Here we remark that the volatility is a free parameter and it is normally estimated by the investors. In a moment we will explain how an investors decide whether to buy an option or no based on the comparisons between the estimated value of the volatility and the historical value appearing on the charts. Understanding this aspect will give us a deeper idea about the importance of doing accurate predictions over the value of the volatility. 

\subsection{The volatility as a parameter for deciding to buy or not an Option}

The history of the Option market is long. It started in 1,972 when the first option was created. The BS formula, able to predict the dynamics of the prices of the Options, was formulated during the same year \cite{Stockbook}. This was the first consistent analytical approach portraying the dynamics of the Option market. The brilliance of Black and Scholes was on the fact that they were able to generate the portfolio free of random fluctuations as we have explained before in eq. (\ref{forever}). It is interesting to see that the BS equation is still widely used by investors in order to decide whether or not it deserves to invest over some specific option related to some stock. The biggest limitation of the BS equation, as we have remarked before, is the fact that the volatility has to be estimated. In other words, there is no way to make predictions about the volatility value inside the BS equation. The role of the volatility over the prices of an Option can be understood if we first explain how the Option market operates. The Option market involves two parties, they are: The holder of the Option and the writer of the same contract \cite{Stockbook}. The Options themselves also come in two different ways, they are: 1). The Call Option, where the holder has the right to buy the shares of a stock at some predetermined price. 2). The Put Option, where the holder has the right to sell the shares of some stock at some predetermined price. In general terms, the Options are contracts between some parties. For the Call Options, the Holder wishes  an increase on the prices of the stock (Bullish behavior), while the writer expects the opposite (Bearish behavior). These expectations occur because for the Call Option, the Holder has the right to buy a stock in future at the same price as the stock is priced today (at the moment of signing the contract).
\begin{figure}
	\centering
\includegraphics[width=0.7\textwidth]{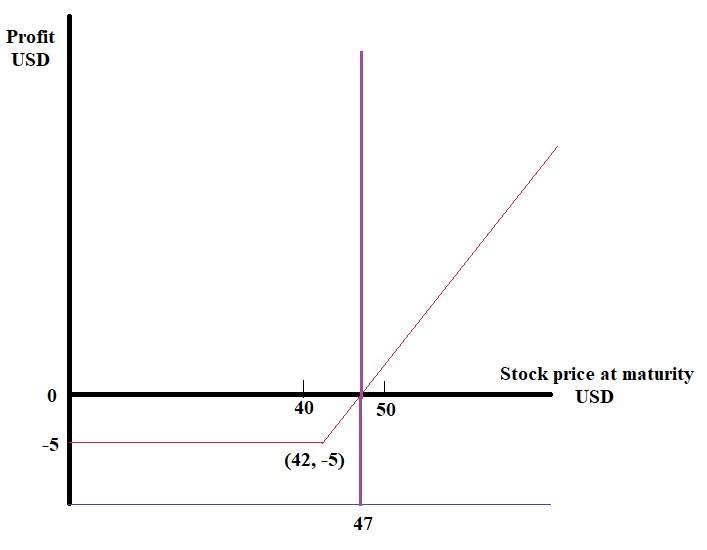}
	\caption{Profits received by the Holder of an European Call Option for an Option price of $5USD$ and a strike price of $42USD$. This is an example of a zero-sum game. Example taken from \cite{Stockbook}}
	\label{Fig.8}
\end{figure}
\begin{figure}
	\centering
\includegraphics[width=0.7\textwidth]{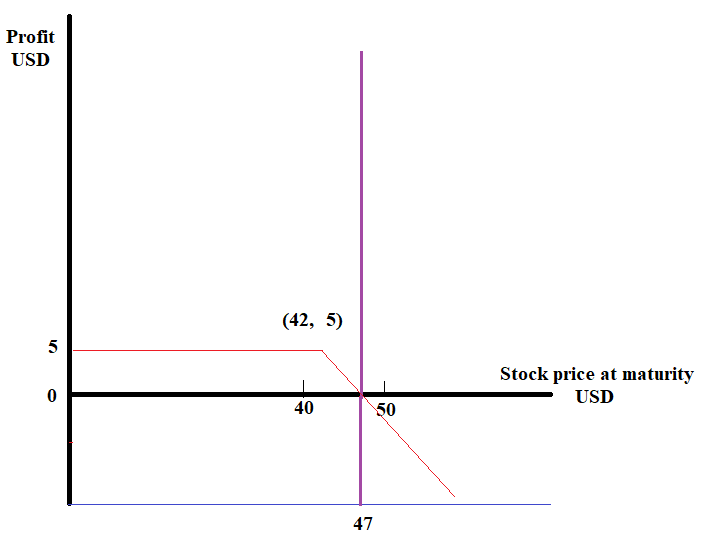}
	\caption{Profits received by the Writer of an European Call Option with an Option price of $5USD$ and a strike price of $42USD$. This is an example of a zero-sum game. Example taken from \cite{Stockbook}}
	\label{Fig.9}
\end{figure}
Let's develop an example as follows. Imagine that the Holder buys a Call Option at $5USD$ per share. This gives him/her the right to buy some specific Stock of some company at the strike price of $42USD$. For the Holder to exercise the Option, he/she needs it to have at least a stock price larger than $47USD$ per share. This is the case because the Holder bought the Option at $5USD$, starting then with a negative balance of $5USD$ per share as it is illustrated in the figure (\ref{Fig.8}). He/she will have this negative balance when he/she does not exercise the Option, which is the case when the stock price is inferior to $47USD$ per share. If by the expiration date of the Option, the price is larger than $47USD$ per share, then the Holder will earn some positive profits after exercising the Option. It is at this point where the writer starts to lose profits as it can be seen from the figure (\ref{Fig.9}), which analyzes the profits of the writer for the same Option and stock price. Note that the Writer earns positive profits if the Option is not exercised by the Holder. In summary, the Holder earns money when he exercises the Option after verifying that the price of the same Option is larger than $47USD$ (for closing prices of the stock between $47USD$ and $42USD$, the European Option could be also exercised in order to minimize the losses suffered by the Holder). What he earns (The Holder) is the difference between the price of the Stock minus what he/she paid for the Option, multiplied by the number of shares involved in the process. On the other hand, the writer makes money when the Holder does not exercise the Option after verifying that the Stock price closes at a value inferior to $47USD$ (more strictly for stock values inferior to $42USD$ because for stock prices between $42USD$ and $47USD$, the Holder could still exercise the European Call Option for minimizing his/her loses). Here we want to remark that while the maximum possible profit for the writer is $5USD$ per share (see the figure (\ref{Fig.9})), the maximum profit for the Holder is just limited by the price of the stock involved in the process.\\
The Put Option is defined as the one where the Holder, once he/she buys it, he/she has the right to sell a stock at some specific pre-arranged price. This means that the Holder of the Option pays to the Writer of the Option some amount in order to get the right to sell a stock at a pre-specified price, which is usually the price of the stock at the day of buying the Option. In this case, the Holder of the Option wants the price of the stock to fall (Bearish), while the writer of the Option, wishes the prices to increase (Bullish). The figure (\ref{Fig.10}), illustrates an example for the possible profits earned by the Holder of an European Call Option. We can notice that the lower the price of the stock is, the larger are the earnings of the Holder.
\begin{figure}
	\centering
\includegraphics[width=0.7\textwidth]{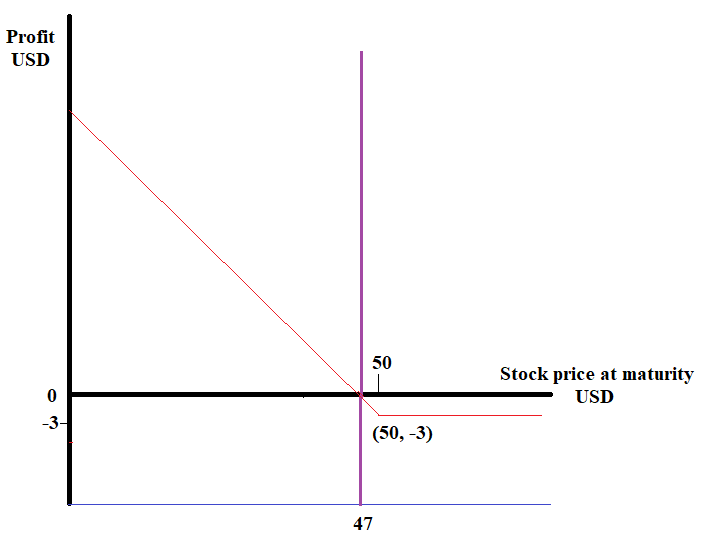}
	\caption{Example of the possible profits received by the Holder of an European Put Option. The Option price is $3USD$ and the strike price is $50USD$. This is an example of a zero-sum game. Example taken from \cite{Stockbook}}
	\label{Fig.10}
\end{figure}
In this particular example, the cost of the Option is $3USD$. Then naturally, the Holder loses $3USD$ per share and then the Writer of the same Option earns $3USD$ per share as it can be seen from the figure (\ref{Fig.11}). 
\begin{figure}
	\centering
\includegraphics[width=0.7\textwidth]{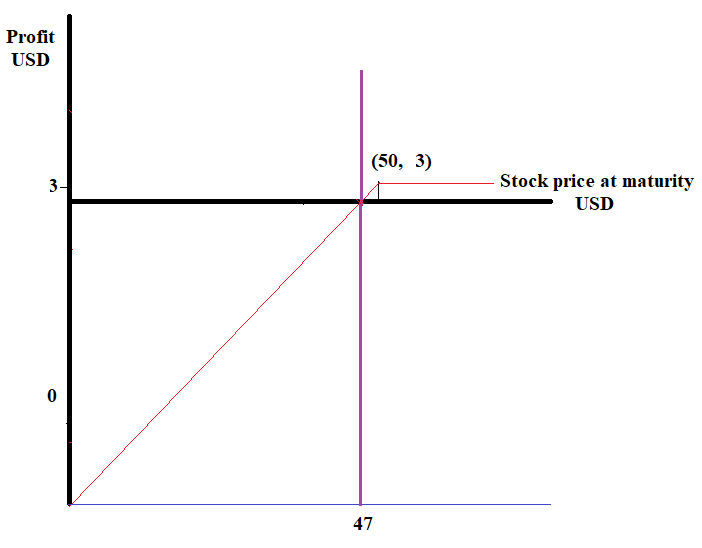}
	\caption{Example of the possible profits received by the Writer of an European Put Option. The Option price is $3USD$ and the strike price is $50USD$. This is an example of a zero-sum game. Figure taken from \cite{Stockbook}}
	\label{Fig.11}
\end{figure}
The figures (\ref{Fig.10}) and (\ref{Fig.11}) are mirror images of each other. This is a typical relation for the pay-offs between two players in a zero-sum game. Inside the European Call Option, for this example, the Holder gets positive pay-offs when the price of the stock is inferior to $47USD$. This is the case because the Holder has the right to sell the stock at $50USD$, no matter what happens with the prices of the stock. Since he/she has to recover the investment of $3USD$, then he/she will only exercise the Option if the price of the involved stock falls below $47USD$ (The Holder could also decide to exercise the European Put Option for prices of the stock between $47USD$ and $50USD$, if by the time of closing, he/she wishes to minimize his/her losses). For prices of the stock larger than this value ($47USD$), the Holder will still possibly exercise if the price of the stock is inferior to $50USD$, just for minimizing losses.
However, if the prices of the stock increase over $50USD$, exercising the Option will bring losses larger than $3USD$ per share to the Holder. Then the Option in this case will never be exercised and then the Writer will earn $3USD$ per share, while the Holder will lose the same amount. For this particular example, the highest pay-off for the Holder is $47USD$, while the Writer also loses the same amount in this situation. Here again, the highest possible pay-off received by the writer is just the price of the Option per share and in the same situation, the Holder would lose the same amount.

\section{The Merton Garman equation}   \label{Merton}

In this section, we consider the MG financial equation, which is the extension of the BS case but this time considering the volatility as a stochastic variable \cite{Op3, Merton2}. When the Option price and the volatility are both stochastic variables, then the market is incomplete \cite{B}. Modelling the volatility is a big challenge and several approaches have been proposed \cite{45}. However, here we consider the approach considered in \cite{B}, which corresponds to the generic situation. We can consider the set of equations

\begin{eqnarray}
\frac{dS}{dt}=\phi Sdt+S\sqrt{V}R_1,\nonumber\\
\frac{dV}{dt}=\lambda+\mu V+\zeta V^\alpha R_2.
\end{eqnarray}
The volatility enters in these equations as the stochastic variable $V=\sigma^2$. $S$ is the stochastic variable representing the price of the security and $\phi$, $\lambda$, $\mu$ and $\zeta$ are constants or better, free-parameters of the system \cite{51}. The variables $R_1$ and $R_2$ are the Gaussian noises, corresponding to each of the variables under analysis. They appear correlated in the following form

\begin{equation}   \label{Correlation}
<R_1(t')R_1(t)>=<R_2(t')R_2(t)>=\delta(t-t')=\frac{1}{\rho}<R_1(t)R_2(t')>.    
\end{equation}
Here $-1\leq\rho\leq1$, and the bra-kets $<AB>$ correspond to the correlation between $A$ and $B$. The fact that the random noises between both variables are correlated, as it appears on eq. (\ref{Correlation}), suggests in advance an important connection between the prices of the stocks and the values taken by the volatility. This is an important aspect of the MG analysis which will be reflected at the moment of using the gauge principle. Let's consider now a function $f$, depending on the Stock price, the time, as well as on the white noises. By using the Ito calculus, it is possible to derive the total derivative in time of this function as follows

\begin{eqnarray}
\frac{df}{dt}=\frac{\partial f}{\partial t}+\phi S\frac{\partial f}{\partial S}+(\lambda+\mu V)\frac{\partial f}{\partial V}+\frac{\sigma^2S^2}{2}\frac{\partial^2f}{\partial S^2}+\rho V^{1/2+\alpha}\zeta\frac{\partial^2f}{\partial S\partial V}+\frac{\zeta^2V^{2\alpha}}{2}\frac{\partial^2f}{\partial V^2}+\sigma S\frac{\partial f}{\partial S}R_1+\zeta V^\alpha\frac{\partial f}{\partial V}R_2.    
\end{eqnarray}
We can now separate the stochastic terms from the non-stochastic ones. The result is

\begin{equation}   \label{HK}
\frac{df}{dt}=\Theta+\Xi R_1+\psi R_2.    
\end{equation}
Here we have defined 
 
\begin{eqnarray}
\Xi=\sigma S\frac{\partial f}{\partial S}, \;\;\;\;\; \;\;\;\;\; \;\;\;\;\;\psi=      
\zeta V^\alpha\frac{\partial f}{\partial V},\nonumber\\
\Theta=\frac{\partial f}{\partial t}+\phi S\frac{\partial f}{\partial S}+(\lambda+\mu V)\frac{\partial f}{\partial V}+\frac{\sigma^2S^2}{2}\frac{\partial^2f}{\partial S^2}+\rho V^{1/2+\alpha}\zeta\frac{\partial^2f}{\partial S\partial V}+\frac{\zeta^2V^{2\alpha}}{2}\frac{\partial^2f}{\partial V^2},
\end{eqnarray} 
The notation used here is the same proposed in \cite{B}. 
 
\subsection{Derivation of the Merton-Garman equation}
 
The derivation of the MG equation, is a very important mathematical exercise. We start by considering two different options, here defined as $C_1$ and $C_2$, on the same underlying security with strike prices and maturities given by $K_1$, $K_2$, $T_1$ and $T_2$ respectively. We can then create a portfolio 
 
 \begin{equation}
\Pi=C_1+\Gamma_1C_2+\Gamma_2S.     
 \end{equation}
By using the result (\ref{HK}), we can define the total derivative of the portfolio with respect to time as
 
 \begin{equation}
\frac{d\Pi}{dt}=\Theta_1+\Gamma_1\Theta_2+\Gamma_2\phi S+\left(\Xi_1+\Gamma_1\Xi_2+\Gamma_2\sigma S\right)R_1+\left(\psi_1+\Gamma_1\psi_2\right)R_2.     
 \end{equation}
This result is a consequence of identifying $f(t)=C_1$ or $f(t)=C_2$ in eq. (\ref{HK}). Although the market is incomplete in this case, we still can create a hedged folio able to satisfy the condition $\frac{d\Pi}{dt}=r\Pi$. The result appears after removing the white noises $R_1$ and $R_2$. This is achieved after imposing the following conditions

\begin{eqnarray}
\psi_1+\Gamma_1\psi_2=0,\nonumber\\
\Xi_1+\Gamma_1\Xi_2+\Gamma_2\sigma S=0.
\end{eqnarray}
We can then solve these equations for $\Gamma_1$ and $\Gamma_2$. We can now define the parameter

\begin{eqnarray}   \label{beta}
\beta(S, V, t, r)=\frac{1}{\partial C_1/\partial V}\left(\frac{\partial C_1}{\partial t}+(\lambda+\mu V)\frac{\partial C_1}{\partial S}+\frac{VS^2}{2}\frac{\partial^2 C_1}{\partial S^2}+\rho V^{1/2+\alpha}\zeta\frac{\partial^2 C_1}{\partial S\partial V}+\frac{\zeta^2V^{2\alpha}}{2}\frac{\partial^2 C_1}{\partial V^2}-rC_1\right)\nonumber\\
=\frac{1}{\partial C_2/\partial V}\left(\frac{\partial C_2}{\partial t}+(\lambda+\mu V)\frac{\partial C_2}{\partial S}+\frac{VS^2}{2}\frac{\partial^2 C_2}{\partial S^2}+\rho V^{1/2+\alpha}\zeta\frac{\partial^2 C_2}{\partial S\partial V}+\frac{\zeta^2V^{2\alpha}}{2}\frac{\partial^2 C_2}{\partial V^2}-rC_2\right)
\end{eqnarray}
The parameter $\beta$ in the MG equation is defined as the market price volatility risk because the higher its value is, the lower is the intention of the investors to risk. Then the risk of the market is always included inside the MG equation. Since the volatility is not traded in the market, then it is not possible to make a direct hedging process over this quantity \cite{B}. In \cite{66}, it was verified that the value of $\beta$ is different from zero. With all these arguments, the MG equation is obtained by re-expressing the equation (\ref{beta}) in the form

\begin{equation}   \label{MGE}
\frac{\partial C}{\partial t}+rS\frac{\partial C}{\partial S}+(\lambda+\mu V)\frac{\partial C}{\partial V}+\frac{1}{2}VS^2\frac{\partial^2 C}{\partial S^2}+\rho\zeta V^{1/2+\alpha}S\frac{\partial^2 C}{\partial S\partial V}+\zeta^2 V^{2\alpha}\frac{\partial^2 C}{\partial V^2}=rC.  
\end{equation}
Here the effects of $\beta$ appear contained inside the modified parameter $\lambda$ in this equation. This can be perceived from the shift $\lambda\to \lambda-\beta$ in eq. (\ref{MGE}).

\subsection{Hamiltonian form of the Merton-Garman equation}

It has been demonstrated before, that after doing the corresponding change of variable, the MG equation takes its Hamiltonian form. The change of variable applies for the prices of the stock and for the volatility as follows

\begin{eqnarray}   \label{newvariables}
S=e^x,\;\;\;\;\;-\infty<x<\infty,\nonumber\\
\sigma^2=V=e^y,\;\;\;\;\;-\infty<y<\infty.
\end{eqnarray}
By using these relations, eq. (\ref{MGE}) becomes \cite{B}

\begin{equation}
\frac{\partial C}{\partial t}+\left(r-\frac{e^y}{2}\right)\frac{\partial C}{\partial x}+\left(\lambda e^{-y}+\mu-\frac{\zeta^2}{2}e^{2y(\alpha-1)}\right)\frac{\partial C}{\partial y}+\frac{e^y}{2}\frac{\partial^2 C}{\partial x^2}+\rho\zeta e^{y(\alpha-1/2)}\frac{\partial^2 C}{\partial x\partial y}+\zeta^2 e^{2y(\alpha-1)}\frac{\partial^2 C}{\partial y^2}=rC.  
\end{equation}
The Schr\"odinger-like equation is

\begin{equation}   \label{SEMG}
\frac{\partial C}{\partial t}=\hat{H}_{MG}C,   
\end{equation}
with the MG Hamiltonian defined as

\begin{equation}  \label{MGHamilton}
\hat{H}_{MG}=-\frac{e^y}{2}\frac{\partial^2 }{\partial x^2}-\left(r-\frac{e^y}{2}\right)\frac{\partial }{\partial x}-\left(\lambda e^{-y}+\mu-\frac{\zeta^2}{2}e^{2y(\alpha-1)}\right)\frac{\partial}{\partial y}-\rho\zeta e^{y(\alpha-1/2)}\frac{\partial^2 }{\partial x\partial y}-\zeta^2 e^{2y(\alpha-1)}\frac{\partial^2 }{\partial y^2}+r.   
\end{equation}
Exact solutions for the MG equation have been found for the case $\alpha=1$ in \cite{B} by using path-integral techniques. The same equation has been solved for the case $\alpha=1/2$ by using standard techniques of differential equations. The MG equation has two degrees of freedom. In the same way as it happens with the BS Hamiltonian, the MG Hamiltonian is also non-Hermitian and then it does not preserve the information, something typical from the stochastic processes.

\section{Symmetries of the Black Scholes Hamiltonian}   \label{Symmetr...}

The symmetries of the Black Scholes Hamiltonian were analyzed in \cite{Equivalence, Ivan}. The symmetries we care about are those related to the changes of the prices of the stocks. We then define the operator $U=e^{\omega \theta(x)}$, which defines the changes of the system with respect to the prices of the stocks. The operator $U$ in this case is non-unitary because the processes in the stock market are basically stochastic and unitarity is not respected at all. We are concerned about the local transformations where $\theta(x)$ is a parameter depending on $x$, which is a function of the prices of the stock as it can be seen from eq. (\ref{newvariables}). With the BS Hamiltonian defined as $\hat{H}_{BS}$, the operator $U$ would be a symmetry of the system if the relation $[\hat{H}_{BS}, U]=0$ were satisfied. However, it has been demonstrated before in \cite{Equivalence}, that $U$ is not a local symmetry of the system represented by the BS equation. This means that the commutator satisfies the relation

\begin{equation}   \label{NESym}
[\hat{H}_{BS}, U]\neq0.
\end{equation}
The transformation represented by $U$, becomes a symmetry when we include a gauge field which comes out to be the stochastic volatility \cite{Equivalence}. Let's analyze this issue in more detail. If we operate with $U$ over the Hamiltonian $\hat{H}_{BS}$, we obtain

\begin{equation}   \label{ModHamiltonia}
\hat{H}_{BS}\to \hat{H}_{BS}+\frac{\sigma^2\omega(1+\omega)}{2}\left(\frac{\partial\theta(x)}{\partial x}\right)^2+\sigma^2\omega\left(\frac{\partial\theta(x)}{\partial x}\right)\frac{\partial}{\partial x}+\omega\left(\frac{1}{2}\sigma^2-r\right)\frac{\partial\theta(x)}{\partial x}.
\end{equation}
Then the local symmetry under changes of the prices of the stocks is not satisfied by the standard BS equation. If we want to restore this symmetry, we need to introduce the stochastic volatility with its corresponding transformation under local changes of the prices, which compensate any variation of the Hamiltonian $\hat{H}_{BS}$ under the same transformations. Since the volatility enters as a gauge field, then the ordinary derivative becomes now a covariant derivative, defined as

\begin{equation}   \label{CovDer}
\frac{\partial}{\partial x}\to \frac{\partial}{\partial x}+\hat{p}_y.  
\end{equation}
While $\hat{p}_x=\frac{\partial}{\partial x}$ is the "momentum" associated to the stock prices, $\hat{p}_y$ is the momentum associated to the stochastic volatility. Then the extended BS Hamiltonian able to satisfy the local symmetry under the changes $U=e^{\omega \theta(x)}$, is defined as 

\begin{equation}   \label{MertonMOdified}
\hat{H}_{BS}\to\hat{H}_{gauge}=\frac{\sigma^2}{2}\left(-\hat{p}_x-\hat{p}_y\right)\left(\hat{p}_x+\hat{p}_y\right)+\left(\frac{1}{2}\sigma^2-r\right)\left(\hat{p}_x+\hat{p}_y\right)+r.
\end{equation}
Here we have defined the extended BS Hamiltonian as the gauge-Hamiltonian as the subindex $gauge$ appearing in eq. (\ref{MertonMOdified}) suggests. It has been proved that this Hamiltonian is a special case of the MG Hamiltonian defined in eq. (\ref{MGHamilton}) \cite{Equivalence}. The Hamiltonian (\ref{MertonMOdified}) can be expressed as 

\begin{equation}   \label{MertonMOdified2}
\hat{H}_{gauge}=-\frac{\sigma^2}{2}\hat{p}_x^2+\left(\frac{1}{2}\sigma^2-r\right)\hat{p}_x-\frac{\sigma^2}{2}\hat{p}_y^2-\sigma^2\hat{p}_x\hat{p}_y+\left(\frac{1}{2}\sigma^2-r\right)\hat{p}_y+r,
\end{equation}
after doing the corresponding expansion. The gauge invariance in guarantee if the following conditions are satisfied

\begin{eqnarray}   \label{gaugevalid}
\left(\frac{\partial\theta}{\partial x}\right)^2=\frac{\omega}{1+\omega}\left(\frac{\partial\theta}{\partial y}\right)^2,\nonumber\\
\left(\frac{\partial\theta}{\partial x}\right)\hat{p}_x=\left(\frac{\partial\theta}{\partial y}\right)\hat{p}_y,\nonumber\\
\frac{\partial\theta}{\partial x}+\frac{\partial\theta}{\partial y}-4\frac{\partial^2\theta}{\partial x\partial y}=\frac{2r}{\sigma^2}\left(\frac{\partial\theta}{\partial x}+\frac{\partial\theta}{\partial y}\right).
\end{eqnarray}
Under these conditions, eq. (\ref{MertonMOdified2}) represents the MG Hamiltonian, when the following relations between parameters are satisfied

\begin{eqnarray}   \label{Volcoeff}
\zeta^2=e^{-2y\left(\alpha-\frac{3}{2}\right)},\nonumber\\
\rho\zeta=e^{-y\left(\alpha-\frac{3}{2}\right)},\nonumber\\
r=\lambda e^{-y}+\mu.
\end{eqnarray}
These relations represent a family of Hamiltonians consistent with the MG one. The relations (\ref{Volcoeff}) reduce the volatility to one additional parameter, unless certain combinations of the free parameters have some stochastic nature. The conditions (\ref{Volcoeff}) are then very restrictive and in this paper we will omit them. Then our Hamiltonian defined in eq. (\ref{MertonMOdified2}), together with the conditions (\ref{gaugevalid}), is not necessarily the MG Hamiltonian but rather a different Hamiltonian which is locally equivalent to the BS Hamiltonian. If we solve the second equation from the group of equations (\ref{gaugevalid}), and we use the first equation from the same group, we get

\begin{equation}   \label{relationvol}
\frac{\hat{p}_x}{\hat{p}_y}=\frac{\left(\frac{\partial\theta}{\partial y}\right)}{\left(\frac{\partial\theta}{\partial x}\right)}=\pm\sqrt{\frac{\omega}{1+\omega}}.   
\end{equation}
For $\omega>>1$, this means that $\hat{p}_x\approx
\pm\hat{p}_y$. No matter which value is taken by the parameter $\omega$, the proportionality between the absolute value of the momentum associated to the changes of prices of stock and the momentum related to changes on the volatility, means that if we consider that the price of an option increases with the volatility, namely, $\hat{p}_y>0$, then the price of the option can increase or decrease with the price of the related stock. We conclude then that the positive sign in eq. (\ref{relationvol}) is ideal for the holders of a Call option while the negative sign is the ideal scenario for the holders of the Put options. This is the case because Call Options are always increasing in price with the price of the stock, while the opposite applies to the Put Options. No matter what option we consider, an increase on the volatility is equivalent to an increase of the price of the option. This is consistent with observations of the patterns of the stock market. Finally, the last relation appearing inside the group of equations (\ref{gaugevalid}), brings out the result 

\begin{equation}   \label{theone}
1+\frac{\hat{p}_x}{\hat{p}_y}-4\frac{\frac{\partial^2\theta}{\partial x\partial y}}{\frac{\partial\theta}{\partial x}}=\frac{2r}{\sigma^2}\left(1+\frac{\hat{p}_x}{\hat{p}_y}\right).
\end{equation}
This result is trivial when $2r=\sigma$, which is precisely the case where the Hamiltonian (\ref{BSHamiltonian}) becomes Hermitian and then the information of the Option market is preserved. Then eq. (\ref{theone}) becomes a way to know the conditions under which the information is not preserved in the stock market. Eq. (\ref{theone}) can be expressed as 

\begin{equation}   \label{oneplus}
1+\frac{\hat{p}_x}{\hat{p}_y}=4\frac{\sigma^2}{(\sigma^2-2r)}\frac{\frac{\partial^2\theta}{\partial x\partial y}}{\frac{\partial\theta}{\partial x}}.
\end{equation}
As we can see, the condition $\sigma^2=2r$ requires $\frac{\partial^2\theta}{\partial x\partial y}=0$ too in order to keep the left-hand side of eq. (\ref{oneplus}) finite.

\subsection{Local equivalence of the Merton Garman and the Black Scholes equation: The gauge principle}

\begin{figure}
	\centering
\includegraphics[width=0.7\textwidth]{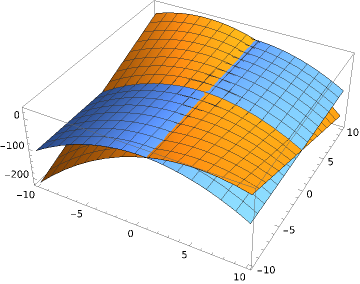}
	\caption{The MG and the BS Hamiltonian. The yellow plot corresponds to the MG case with non-trivial volatility, while the blue curve is the BS equation, considering the volatility as a parameter (constant for the purpose of the figure).}
	\label{MG1}
\end{figure}

For the family of MG Hamiltonians satisfying the local symmetry under changes of the prices, namely, the Hamiltonian (\ref{MertonMOdified2}); the volatility, being a gauge field, is the responsible of restoring the symmetry of the system under local changes of the prices. The gauge principle suggests then that the family of MG Hamiltonians consistent with the gauge principle, is locally equivalent to the BS Hamiltonian. In other words, the MG equation consistent with the gauge principle is locally equivalent to the BS equation. This means that if we switch-off the momentum associated to the volatility $\hat{p}_y$, we will then recover the BS equation. This is a trivial task easy to verify. The key point of this section is the interpretation behind switching off the momentum associated to the stochastic volatility. The condition to analyze is 

\begin{equation}
\hat{H}_{gauge}=\hat{H}_{MG}^{\hat{p}_yC\to0}\to\hat{H}_{BS}. 
\end{equation}
The condition $\hat{p}_yC\to0$ means that any variations of the prices of the options due to changes of the volatility are completely ignored when we consider the the BS limit. The graphic representation of this situation can be perceived on the figure (\ref{MG1}). We can perceive the volatility as a field generating some "weight" over the the option prices. This weight deviates the prices of the option from the usual path they would take if the volatility is suppressed. When the momentum associated to the volatility vanishes, then from the figure we see that the BS Hamiltonian (blue plot) intersects with the MG Hamiltonian (yellow figure) because in this limit both Hamiltonians are the same.
\begin{figure}
	\centering
\includegraphics[width=0.7\textwidth]{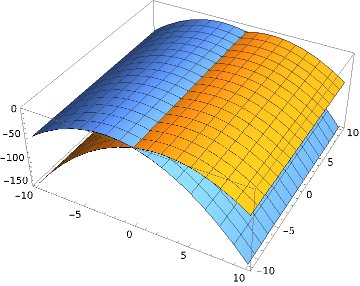}
	\caption{The BS Hamiltonian with the volatility appearing as a free-parameter. The blue plot corresponds to the case with $\frac{1}{2}\sigma^2<r$, while the yellow plot corresponds to the case with $\frac{1}{2}\sigma^2>r$.}
	\label{MG2}
\end{figure}
From the same figure it can be seen that in general, the MG Hamiltonian does not match the BS Hamiltonian. The equality occurs when $\hat{p}_y=0$ as it has been mentioned, but also for the non-trivial case where $\hat{p}_y\neq0$ as far as the eigenvalues of $\hat{p}_x$ satisfy

\begin{equation}   \label{Surprise}
\hat{p}_x\vert C>+\frac{1}{2}\hat{p}_y\vert C>=\left(\frac{1}{2}-\frac{r}{\sigma^2}\right)\vert C>.
\end{equation}
Note that this relation becomes trivial on the right-hand side when $\sigma^2=2r$, which is precisely the condition for the Option market to preserve the information based on the Hermiticity condition of the BS Hamiltonian. When this occurs, we then get the relation

\begin{equation}   \label{sur2}
2\hat{p}_x=-\hat{p}_y,     
\end{equation}
which is only valid when an increasing value on the price of an Option is related to a decreasing value of the price of the corresponding stock (Bearish behavior). The result (\ref{Surprise}), is surprising because this means that even when there is a non-trivial value of the stochastic volatility (taken as a variable), the BS Hamiltonian can be recovered from the gauge Hamiltonian as far as the combination $\hat{p}_x+\frac{1}{2}\hat{p}_y$ obeys the the equality (\ref{Surprise}).

\section{The Martingale condition with stochastic volatility}   \label{Martingalelalala}

In \cite{Equivalence, Ivan}, it was proved that some relations between the field related to the prices of the stock and the field corresponding to the stochastic volatility emerge when we consider the vacuum or martingale condition inside the MG scenario. The martingale was then defined as 

\begin{equation}   \label{newmartin}
\hat{H}_{MG}e^{x+y}=\hat{H}_{MG}S(x, y, t)=0.    
\end{equation}
For a non-trivial behavior of the stochastic volatility ($\hat{p}_y\neq0$), this condition corresponds to a martingale condition if the result

\begin{equation}   \label{corona4}
\lambda+e^y\left(\mu+\frac{\zeta^2}{2}e^{2y(\alpha-1)}+\rho\zeta e^{y(\alpha-1/2)}\right)=0,
\end{equation}
is satisfied. It was demonstrated in \cite{Equivalence, Ivan} that the combination of eqns. (\ref{Volcoeff}) and (\ref{corona4}), brings out the simple relation 

\begin{equation}   \label{THisonemama}
e^{2y}+\mu e^y+\lambda=0,    
\end{equation}
which can be easily solved, finding some specific relation between the prices of the volatility and some of the free-parameters ($\mu$ and $\lambda$) of the system. The relation (\ref{THisonemama}) is valid at the equilibrium (martingale) state of the system because the its origin comes from eq. (\ref{newmartin}). The same relation suggests that at the equilibrium condition, the volatility depends on some free-parameters of the MG Hamiltonian. Outside the equilibrium, naturally, the condition (\ref{THisonemama}) is not necessarily satisfied. Now we can analyze the martingale condition for the gauge Hamiltonian defined in eq. (\ref{MertonMOdified2}), without the restrictions of the MG case. In other words, we will now generalize the situation for cases where it is also possible to have $\hat{H}_{gauge}\neq\hat{H}_{MG}$. Here we remark that $\hat{H}_{gauge}$ covers a more general case in comparison with the restrictive MG Hamiltonian. From the Hamiltonian (\ref{MertonMOdified2}), we can define the Martingale condition as 

\begin{equation}
\hat{H}_{gauge}\vert S>=0.    
\end{equation}
Let's take the martingale state as a series expansion $\vert S>=\sum_n\phi^n=\sum_n(\phi_x\phi_y)^n$, as it was suggested in \cite{Ivan}. In such a case, it is possible to demonstrate that the martingale condition gives the result 

\begin{equation}   \label{FullMatringale}
\phi_{xvac}=\phi_{rvac}.    
\end{equation}
This simply suggests that the martingale condition emerges when the variations of the prices of the option, due to changes in the prices of the stock, equals the variations of the prices of the options due to changes in the volatility. This means that the market equilibrium can be achieved with non-trivial values of the volatility, as far as the condition (\ref{FullMatringale}) is satisfied. This surprising result is a natural consequence of the gauge principle. Note that this condition was obtained before when $\omega>>1$ in eq. (\ref{relationvol}). Then the gauge principle suggests important relational connections between the prices of the stocks and the values associated to the volatility of the market. 

\section{Conclusions}   \label{Conclusions}

In this paper we have demonstrated the power of the gauge principle when it is applied to the option market through the BS Hamiltonian. First, the gauge principle shows that the MG equation is locally equivalent to the BS equation. However, the Hamiltonian obtained by imposing the gauge symmetry (gauge Hamiltonian) under changes of prices, is more general than the MG Hamiltonian and it brings out remarkable results. It suggests for example that there is a martingale condition when the variations of the prices of the options with respect to the changes in prices of stocks, are equal to the changes of the prices of the options with respect to changes in volatility. The gauge principle also suggests that the gauge Hamiltonian brings out the same results of the BS Hamiltonian when either of the following two conditions are satisfied, they are: 1). The momentum associated to the volatility is zero. 2). When some specific combination of the momentum associated to the stock prices and the momentum associated to the volatility is satisfied, as eq. (\ref{Surprise}) suggests. Future analysis involving the importance of the gauge principle on the Option market are under analysis. This paper demonstrates that the concept of symmetry and gauge theories, is not only important in physics, where it brings out several important results \cite{physics1, physics2, physics3, physics4, physics5, Mypapers}, but it is also important for understanding the most fundamental principles behind the Option market \cite{Equivalence, Ivan}.

{\bf Acknowledgements}
The author would like to thank Prof. Taksu Cheon from Kochi University of Technology for the hospitality and support during the research visit and presentation carried out at this institution in July 2024. Special thanks to Prof. Taksu Cheon who encouraged the writing of this paper.

\bibliographystyle{unsrt}  


\begin{thebibliography}{1}




\bibitem{BS}
Black, F. and Scholes, M., {\it The pricing of Options and Corporate Liabilities}, Journ. Pol. Ec. {\bf 81} (1973):637.

\bibitem{Vol}
F. de Weert, {\it An Introduction to Options Trading}, John Wiley and Sons, LTD, (2006), ISBN-13 978-0-470-02970-1 (PB), ISBN-10 0-470-02970-6 (PB).

\bibitem{Op3} Merton, R. C., {\it The theory of rational Option Pricing}, Bell Journ. Econ. Manag. Sc. {\bf 4} (1973):141-183.

\bibitem{Merton2} Merton, R. C., {\it Option pricing when underlying stock returns are discontinous}, Journ. Fin. Econ. {\bf 3}, (1976):125.

\bibitem{Equivalence}
Arraut, I., {\it Gauge symmetries and the Higgs mechanism in Quantum Finance}, 2023 EPL {\bf 143} 42001, arXiv:2306.03237 [q-fin.GN].

\bibitem{Ivan}
Arraut, I., Au, A. and Ching-biu Tse, A., {\it Spontaneous symmetry breaking in quantum finance}, EPL {\bf 131} (2020) 6, 68003.

\bibitem{Gauge1}
Carrol S., {\it Spacetime and Geometry:An Introduction to General Relativity}, San Francisco: Addison-Wesley, ISBN 978-0-8053-8732-2.

\bibitem{Gauge2}
Peskin M. E and Schroeder D. V., An introduction to
Quantum Field Theory, CRC press, Taylor and Francis
group, 6000Broken Sound Parkway, NW Suite 300, Boca
Raton Fl. 33487-2742, (2018).

\bibitem{physics1}
Glashow, S.; {\it The renormalizability of vector meson interactions}, Nucl. Phys. 10, 107, (1959).

\bibitem{physics2}
Salam, A. and Ward, J. C., {\it Weak and electromagnetic interactions}, Nuovo Cimento. {\bf 11} (4): 568–577, (1959).

\bibitem{physics3}
Weinberg, S.; {\it A Model of Leptons}, Phys. Rev. Lett. {\bf 19} (21): 1264–66, (1957).

\bibitem{physics4}
Arraut, I.; {\it The origin of the mass of the Nambu–Goldstone bosons}, Int.J.Mod.Phys.A {\bf 33} (2018) 07, 1850041.

\bibitem{physics5}
Arraut, I.; {\it The Nambu-Goldstone theorem in non-relativistic systems}, Int.J.Mod.Phys.A {\bf 32} (2017) 1750127.

\bibitem{Mypapers}
Arraut, I.; {\it The Quantum Yang Baxter conditions: The fundamental relations behind the Nambu-Goldstone theorem}, Symmetry {\bf 11} (2019) 803.

\bibitem{B}
Baaquie, B. E., {\it Quantum Finance: Path integrals and Hamiltonians for options and interestrates}, Cambridge University Press (2004), pp 52-75.

\bibitem{Schro}
Griffiths, D. J., {\it Introduction to Quantum Mechanics}, Prentice Hall, (1995), Upper Saddle River, NJ 07458.

\bibitem{Stockbook}
F. de Weert, {\it An Introduction to Options Trading}, John Wiley and Sons, LTD, (2006), ISBN-13 978-0-470-02970-1 (PB), ISBN-10 0-470-02970-6 (PB).

\bibitem{45} 
Heston, S. L., {\it A Closed-Form Solution for Options with Stochastic Volatility with Application to Bond and Currency Options}, The Review of Financial Studies, 6 (1993): 327.

\bibitem{51}
Baldwin, R. and Taglioni, D. (2007), {\it Trade effects of the euro: A comparison of estimators}, Journ. Econ. Int. {\bf 22} (4): 780–818. 















\end{thebibliography}

\end{document}